\documentclass[aps,prb,superscriptaddress,preprint,endfloats]{revtex4-1}
\usepackage{graphicx,color,bm,amsmath,natbib,url,appendix}

\begin{document}

%\graphicspath{{../../Figures/Dichotomy/}}

\newcommand{\mgb}{MgB$_2$}
\newcommand{\Tc}{T_{\text{c}}}
\newcommand{\Hcii}{H_{\text{c2}}}
\newcommand{\Hac}{H_{\text{ac}}}
\newcommand{\Hdc}{H_{\text{dc}}}
\newcommand{\fms}{f_{\text{MS}}}

\title{Non-equilibrium structural phase transitions of the vortex lattice in {\mgb}}

\author{E.~R.~Louden}
\author{C.~Rastovski} 
\affiliation{Department of Physics, University of Notre Dame, Notre Dame, Indiana 46656, USA}

\author{L.~DeBeer-Schmitt}
\affiliation{Large Scale Structures Group, Neutron Sciences Directorate, Oak Ridge National Laboratory, Oak Ridge, Tennessee 37831, USA}

\author{C.~D.~Dewhurst}
\affiliation{Institut Laue-Langevin, 71 avenue des Martyrs, CS 20156, F-38042 Grenoble cedex 9, France}

\author{N.~D.~Zhigadlo}
\affiliation{Laboratory for Solid State Physics, ETH, CH-8093 Zurich, Switzerland}
\affiliation{Department of Chemistry and Biochemistry, University of Bern, CH-3012 Bern, Switzerland}

\author{M.~R.~Eskildsen}
\altaffiliation{Corresponding author: eskildsen@nd.edu}
\affiliation{Department of Physics, University of Notre Dame, Notre Dame, Indiana 46656, USA}

\date{\today}

\begin{abstract}
We have studied non-equilibrium phase transitions in the vortex lattice in superconducting {\mgb}, where  metastable states are observed in connection with an intrinsically continuous rotation transition.
Using small-angle neutron scattering and a stop-motion technique, we investigated the manner in which the metastable vortex lattice returns to the equilibrium state under the influence of an ac magnetic field.
This shows a qualitative difference between the supercooled case which undergoes a discontinuous transition, and the superheated case where the transition to the equilibrium state is continuous.
In both cases the transition may be described by an an activated process, with an activation barrier that increases as the metastable state is suppressed, as previously reported for the supercooled vortex lattice [E.~R.~Louden {\em et al.}, Phys. Rev. B {\bf 99}, 060502(R) (2019)].
Separate preparations of superheated metastable vortex lattices with different domain populations showed an identical transition towards the equilibrium state.
This provides further evidence that the vortex lattice metastability, and the kinetics associated with the transition to the equilibrium state, is governed by nucleation and growth of domains and the associated domain boundaries.
\end{abstract}

\maketitle

% INTRODUCTION	
\section{Introduction}
For physical systems in equilibrium, it is customary to classify phase transitions as discontinuous (first order) or continuous (second order).
However, the characteristics of non-equilibrium phase transitions may differ significantly.\cite{Henkel:2008wn,Henkel:2010wn}
Vortex matter in type-II superconductors\cite{Blatter:1994gz,Giamarchi:1995tq,Huebener:2001we} offers a conceptually simple two-dimensional system to explore fundamental problems such as non-equilibrium phase transitions and kinetics, structure formation and transformation at the mesoscopic scale, and metastable states.
In addition, vortex matter shows many similarities with magnetic skyrmions;\cite{Pfleiderer_SkL_rev,Nagaosa:2013cc} soft matter systems such as liquid crystals, colloids and granular materials;\cite{Nagel:2017fe} and glasses.\cite{Lubchenko:2007cc}
Insights gained from vortex studies may therfore be applicable to a wide range of material systems.

In the hexagonal two-band superconductor {\mgb} the equilibrium vortex lattice (VL) phase diagram consists of three likewise hexagonal phases connected by a continuous rotation transition.\cite{RefWorks:19,Hirano:2013jx}
Cooling or warming across the equilibrium phase boundaries leaves the VL in robust metastable states.\cite{Das:2012cf}
The metastability is a collective vortex phenomenon most likely due to the presence of robust VL domain boundaries, and is notably not due to pinning.\cite{Rastovski:2013ff}
Field/temperature history dependent metastability has also been observed in connection with structural transitions of the skyrmion lattice (SkL).\cite{Makino:2017hh,Nakajima:2017uc,Bannenberg:2017ws}
Compared to the SkL, the VL can more easily be perturbed by varying the magnetic field and used to study transitions between metastable and equilibrium phases. 
In addition to the SkL, one may expect similarities between the VL and other physical systems governed by domain nucleation and growth, such as martensitic phase transitions~\cite{Wang:2017jw} or domain switching in ferroelectrics.\cite{Shin:2007gu}

\begin{figure*}
    \includegraphics{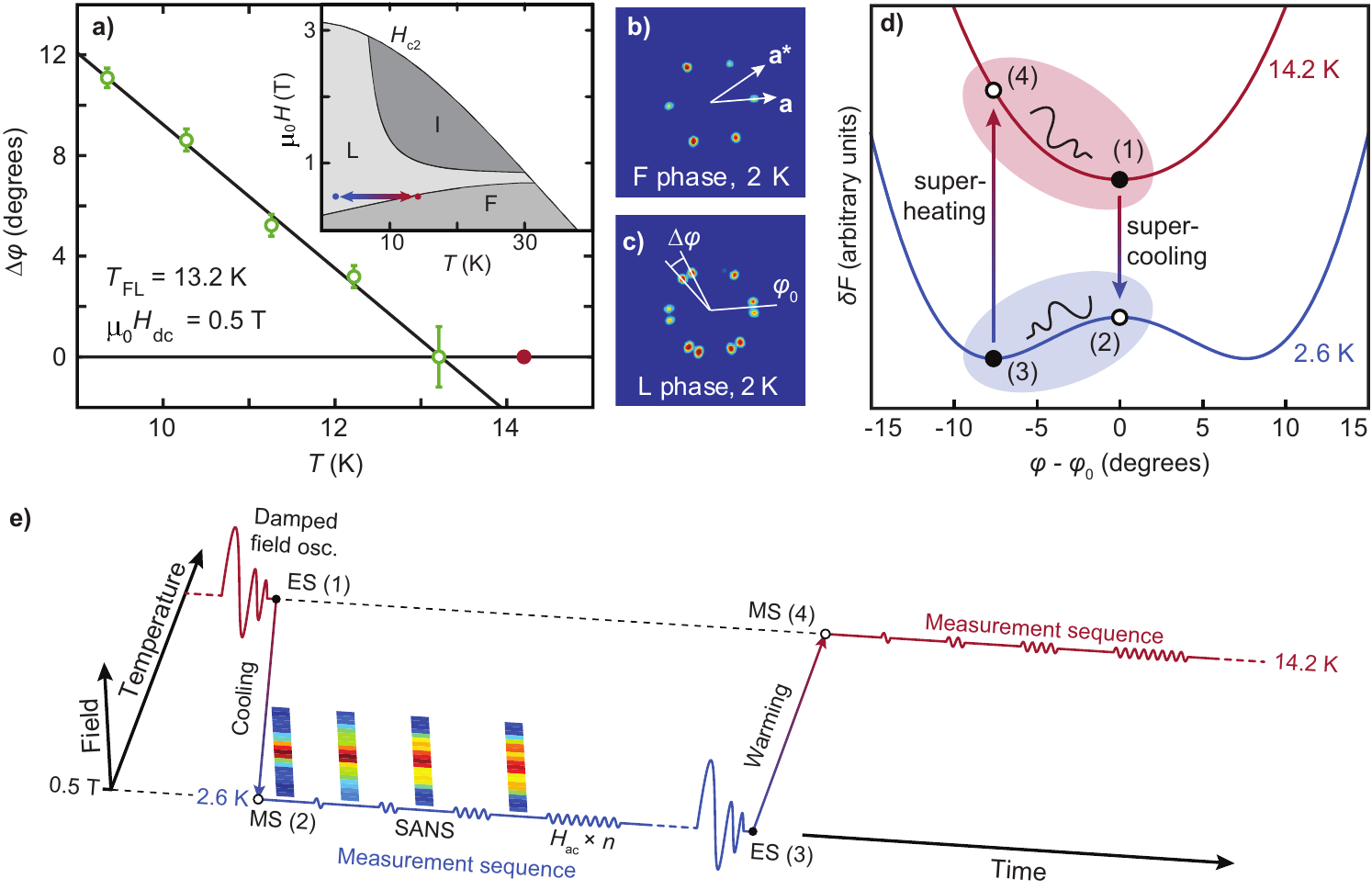}
    \caption{(Color online)
        Vortex lattice configurations for {\mgb}.
        (a) Equilibrium VL Bragg peak separation at 0.5~T (open circles).
        The solid circle indicates the temperature used for measurements of the superheated VL.
        The inset shows the equilibrium VL phase diagram for {\mgb} consisting of three hexagonal configurations.
        SANS diffraction patterns recorded at 0.5~T and 2~K show (b) a MS and (c) an ES VL.
        High symmetry directions within the {\mgb} hexagonal crystal basal plane are shown in (b), and the reference angle ($\varphi_0) $ and Bragg peak splitting ($\Delta \varphi$) in (c).
        (d) Schematic single domain VL free energy curves corresponding to the supercooled and superheated VL configurations.
        Solid (open) circles represent an ES (MS) VL.
        (e) Illustration of SANS measurements; the numbers in parentheses correspond to the same states as in panel (d). Individual colorbars represent the collection of SANS data, each providing one ``slice'' of the full transition sequence in Fig.~\ref{Fig2}(d).
        \label{Fig1}}
\end{figure*}

% Equilibrium Phase Diagram
The equilibrium VL phase diagram for {\mgb} with the magnetic field applied parallel to the $c$ axis is shown in the inset to Fig.~\ref{Fig1}(a).
In both the F and I phases a single global orientational order is observed by small-angle neutron scattering (SANS), indicated by six VL Bragg peaks aligned with respectively the $\bm{a}$ and $\bm{a^*}$ direction within the basal plane.\cite{Das:2012cf}
In the intermediate L phase, the VL rotates continuously from the $\bm{a}$ to the $\bm{a^*}$ orientation,
where the presence of both clockwise and counterclockwise domain rotations leads to 12 Bragg peaks.
A diffraction pattern corresponding to a metastable VL, created by supercooling across the F-L phase boundary, is shown in Fig.~\ref{Fig1}(b).
The corresponding equilibrium VL, obtained at the same place in the phase diagram, is shown in Fig.~\ref{Fig1}(c).
The latter was achieved by applying a damped dc field oscillation, as thermal excitations are insufficient to drive the VL to the equilibrium configuration.\cite{Das:2012cf}
The experimentally determined equilibrium phase diagram has been corroborated by numerical calculations.\cite{Hirano:2013jx}

The simplest model for the single domain VL free energy that allows for a continuous rotation connecting the F, L and I phases is given by
\begin{equation}
    \delta F = K_6 \cos \, [6 (\varphi - \varphi_0)] + K_{12} \cos \, [12 (\varphi - \varphi_0)],
    \label{FreeEnergy}
\end{equation}
where $K_{6/12}$ are field and temperature dependent coefficients.\cite{Zhitomirsky:2004jq,Das:2012cf}
This yields the curves in Fig.~\ref{Fig1}(d), which show a qualitative difference between the metastable F (2.6~K) and L (14.2~K) states.
In the first case (pt.~2), the supercooled F phase is in unstable equilibrium [$d(\delta F)/d\varphi = 0; d^2(\delta F)/d^2\varphi < 0$].
In the second case (pt.~4), the superheated L phase is in a true non-equilibrium configuration [$d(\delta F)/d\varphi \neq 0$].

The difference in the free energy configuration between the supercooled and superheated VL provides the motivation for the current work.
Recent studies of the transition kinetics, associated with driving a supercooled VL from the metastable state (MS) to the equilibrium state (ES) by inducing vortex motion, showed an activated behavior.\cite{Louden:2019bq}
Furthermore, the activation barrier was found to increase as the metastable state was suppressed, corresponding to an aging of the VL where the ac field amplitude and cycle count are equivalent to, respectively, an effective ``temperature'' and ``time''.
Here we report on SANS studies of the VL that compare the MS to ES transition for the supercooled and superheated cases.

% EXPERIMENTAL DETAILS
\section{Experimental Details}
\label{ExpDet}

% SANS details
Measurements were performed on the CG2 General Purpose SANS beam line at the High Flux Isotope Reactor at Oak Ridge National Laboratory, and the D33 beam line at Institut Laue-Langevin. 
The data presented here were collected at D33,\cite{5-42-420} but consistent results were found at both facilities.
All experiments were conducted with the incoming neutrons parallel to the applied magnetic field,\cite{Eskildsen:2011jp} using a tightly collimated beam in order to resolve closely spaced VL Bragg reflections. 
The D33 beam collimation was defined by a 2~mm sample aperture which is comparable to the sample size, and a 10~mm  source aperture separated by 12.8~m.
%(a 20~mm source aperture was used for the initial set-up and alignment).
The azimuthal resolution $w_{\text{res}} = 3.1^{\circ}$ was estimated from the width of the undiffracted beam on the detector, $\Delta q$.
Considering a peak of this size at the expected $q_{\text{VL}} = 0.105$~nm$^{-1}$ for a vortex lattice at 0.5~T, the angular azimuthal resolution is given by
$\tan \left( w_{\text{res}}/2 \right) = \tfrac{1}{2} \Delta q/q_{\text{VL}}$.
The D33 data was collected using a wavelength of $\lambda = 0.7$~nm and spread of $\Delta \lambda / \lambda = 10\%$. 

% Sample Details
We used the same 200~$\mu$g single crystal of {\mgb} ($\Tc = 38$~K, $\mu_0 \Hcii = 3.1$~T) as in prior studies.
The {\mgb} crystal had a flat plate morphology, with an area of $\sim 1 \times 1$~mm$^2$ and a thickness of $\sim 50~\mu$m.
The sample was grown with isotopically enriched $^{11}$B to decrease neutron absorption, using a high pressure cubic anvil technique that has been shown to produce good quality single crystals with a mosaicity of a few tenths of a degree.\cite{Karpinski:2003fe,Kazakov:2005ipb}
The observation of VL diffraction peaks belonging to a single F or I phase at low/high fields excludes the possibility of a polycrystal. 
Demagnetization effects are expected to be negligible for the measurement field and geometry, which is supported by the equality of the applied field ($\mu_0 H$) and measured internal magnetic induction ($B$).\cite{Rastovski:2013ff}
The VL metastability has been confirmed in other single crystals of {\mgb},\cite{Das:2012cf} and recently also in crystals from a different source.\cite{ManniXtal}

% ac Details
Vortex motion was induced using a bespoke coil to apply a controlled number of ac field cycles
parallel to the dc field used to create the VL.
A sinusoidal wave function was used, with peak-to-peak amplitudes between $0.5$ and $1.5$~mT and a frequency of 250~Hz.
The ac amplitudes are roughly two orders of magnitude smaller than that of the damped field oscillation used to obtain the ES VL.
This leads to a gradual evolution of the VL from the MS to the ES, and allows for a detailed study of the relaxation process.
The low frequency is equivalent to a ``fast dc'' field oscillation, but more precise than what can be achieved using the superconducting cryomagnet used to apply the static 0.5~T field.
The ac field frequency and amplitudes were chosen to allow, in a reasonable amount of time, a controlled relaxation of the VL affecting the entire sample volume.
%Increasing the ac amplitude would result in greater vortex motion, requiring fewer cycles to reach the ES and thus reducing the precision.

% RESULTS
\section{Results}

% Equilibrium F-L phase boundary
\subsection{Equilibrium F-L phase boundary}
Prior to beginning systematic measurements of the MS to ES transition, the exact location of the F-L phase boundary with $\textbf{H} \! \parallel \! \textbf{c}$ at $0.5$~T was determined from the VL peak separation ($\Delta \varphi$), as shown in Fig.~\ref{Fig1}(a).
The azimuthal position ($\varphi$) of the VL Bragg peaks are measured relative to the crystalline $\bm{a}$ direction ($\varphi_0$).
At each temperature, a damped dc field oscillation with an initial amplitude of 50~mT around its final value of $0.5$~T was applied to obtain the equilibrium VL prior to the SANS measurements.
Since the VL density is directly proportional to the applied field, this gives rise to a breathing motion where vortices are pushed into and out of the sample.
In superconductors with low pinning, such as {\mgb}, this is expected to result in a well-ordered, equilibrium VL configuration.
We will discuss  this is further detail in Sect.~\ref{ESVL}.

Gaussian multi-peak fits to the data were used to determine $\Delta \varphi$ around $\varphi_0$, with $\Delta \varphi = 0 ^{\circ}$ corresponding to the F phase.
The larger error bars at $13.2$~K result from forcing a two-peak guassian fit to data where the peaks have minimal, if any, separation.
The value of $T_{\text{FL}} = 13.2$~K is consistent with the phase diagram originally established by Das {\em et. al.}\cite{Das:2012cf}
At $14.2$~K, a single peak fit to the data yields a full width half maximum (FWHM) that agrees within error with those obtained for the ES L phase peaks at 2.6~K.

%  Nature of the metastable to equilibrium state transition
\subsection{Nature of the metastable to equilibrium state transition}
% ES Prep
All SANS measurement sequences were performed with a 0.5~T dc field parallel to the crystal $c$ axis.
A schematic illustrating the measurements are shown in Fig.~\ref{Fig1}(e).
Prior to each measurement sequence, a pristine supercooled (superheated) MS VL was prepared:
First, an equilibrium VL was obtained at 14.2~K (2.6~K) by performing a damped oscillation of the dc magnetic field.
Second, the ES VL was cooled (warmed) to 2.6~K (14.2~K) across the F-L phase boundary to obtain a MS.
The temperatures were chosen to lie well within the relevant F or L phase, as shown in Fig.~\ref{Fig1}(a).
As the VL relaxation itself is not thermal, it is not expected to depend on the rate of cooling/warming.
For the measurement sequences we used a stop-motion technique, alternating between imaging the VL by SANS and application of ac field cycles. As the vortex density is directly proportional to the applied field the ac cycles gives rise to a breathing motion with vortices being pushed into and out of the sample, which in turn causes the VL to evolve gradually towards the ES.

\begin{figure*}
    \includegraphics{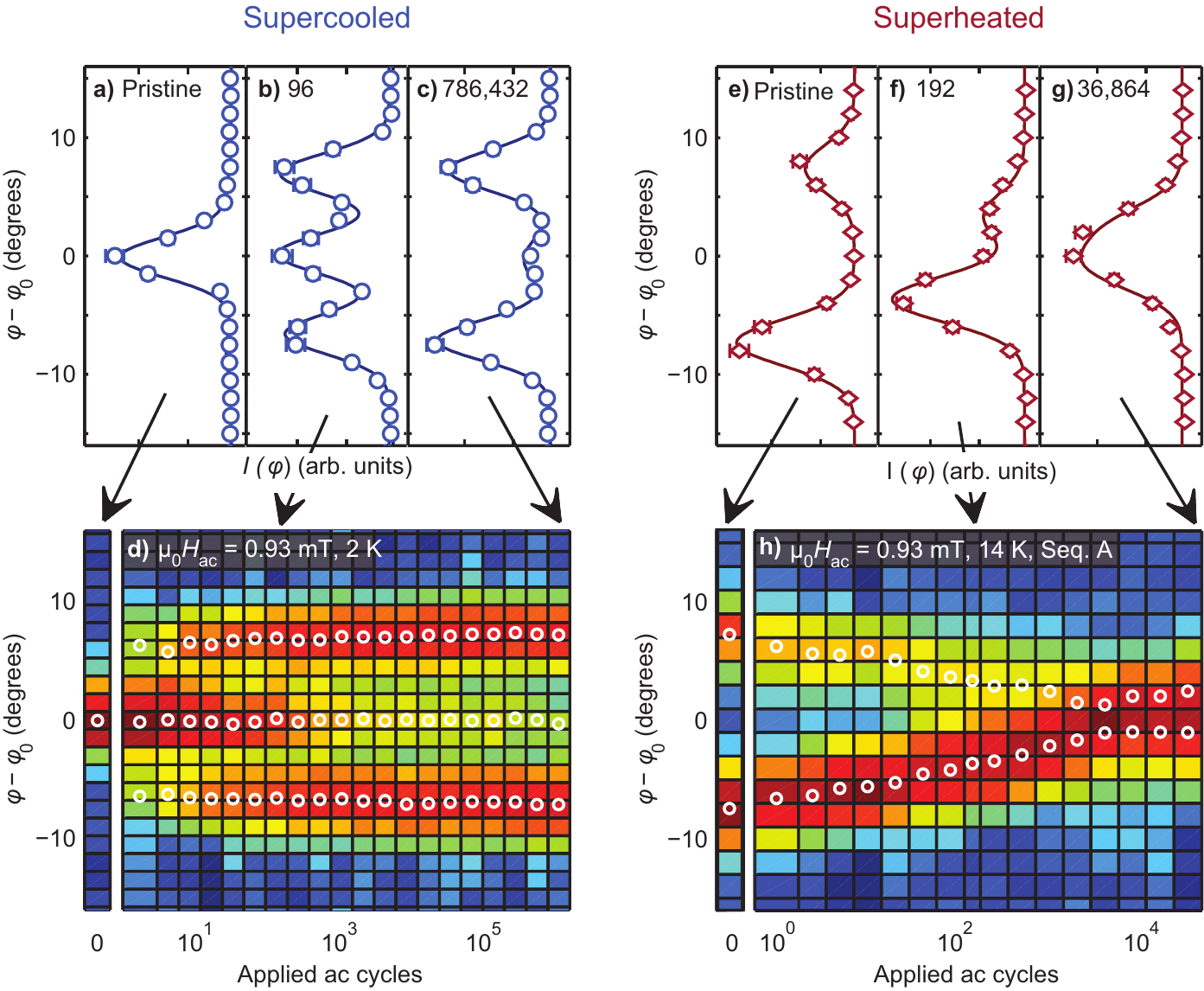}
    \caption{(Color online)
        VL evolution for the MS to ES transitions at 2.6~K and 14.2~K, with $\mu_0 \Hac = 0.93$~mT.
        The supercooled case is summarized in panels (a-d), and the superheated case in panels (e-h).
        Plots of $I(\varphi)$ are shown for three different VL configurations in each case, corresponding to the pristine MS (a,e); a representative intermediate distribution (b,f); and the final measurement (c,g).
        Angles are measured relative to the crystalline $\bm{a}$ axis ($\varphi_0$), as defined in Fig.~\ref{Fig1}(c).
        Colormaps (d,h) show the azimuthal intensity vs.~the number of applied ac cycles; the left colorbar indicates the pristine MS VL. Open circles represent the peak positions obtained by Gaussian multi-peak fits to the data.		
        \label{Fig2}}
\end{figure*}

% 2K Measurement Sequence
A typical measurement sequence for the supercooled MS is illustrated in Fig.~\ref{Fig2}(a) to (d) for $\mu_0 \Hac = 0.93$~mT.\cite{Louden:2019bq}
Panel (a) shows the azimuthal intensity distribution, $I(\varphi)$, for the pristine MS VL with a single Bragg peak at $\varphi = \varphi_0$ corresponding to the F phase.
After applying 96 ac field cycles (b) additional Bragg peaks, corresponding to the L phase, are observed at $\varphi \approx \varphi_0 \pm 7.2^{\circ}$.
This indicates the coexistence of MS F and ES L phase VL domains in the sample.
Following a total of 786,432 cycles (c), the VL has been driven to the ES within most of the sample.
The evolution from the supercooled MS to the ES is summarized in Fig.~\ref{Fig2}(d) and a movie of the transition is included in the Supplemental Material.\cite{SM}
This shows that the ES domains nucleate in their final orientations and grow at the expense of the MS domains.

% 14K Measurement Sequence
An analogous measurement sequence for the superheated MS is shown in Fig.~\ref{Fig2}(e) to (h).
Here, the metastable VL (e) corresponds to the L phase, with two domain orientations at $\varphi \approx \varphi_0 \pm 7.3^{\circ}$.
After 192 ac cycles (f), the Bragg peak separation has decreased roughly by a factor of two.
Following a total of 36,864 cycles (g), the separation is further reduced, resulting in a single, broadened peak. The evolution from the superheated MS to the ES is summarized in Fig.~\ref{Fig2}(h) and in a movie in the Supplemental Material.\cite{SM}
The MS domains rotate continuously towards the final ES position, but never fully merge at $\varphi = \varphi_0$.
This continuous rotation is in sharp contrast to the transition for the supercooled case.
Changing the ac field amplitude affects how quickly the VL returns to the equilibrium configuration but not the qualitative difference between the supercooled and superheated case.
Finally, the data in Fig.~\ref{Fig2}(g) and (h) shows that the ac cycles affect the VL throughout the entire sample as no measurable intensity persists at the angular positions corresponding to the intial MS.
	
% Analysis
To analyze the neutron scattering data, the intensity was binned along the azimuthal direction and fitted with multi-peak gaussians, as shown in Fig.~\ref{Fig2}(a) to (c) and (e) to (g).
The width of the VL Bragg peaks remained constant within our measurement precision,
and a single average value ($\simeq 4^{\circ}$) was used for all gaussian peaks within the same measurement sequence.
The exact value of the width for a given sequence was determined from initial unconstrained fits to the individual SANS measurements where all peaks were clearly resolved.
Bragg peak positions, obtained by multi-peak Gaussian fits to the data, are indicated by the white circles in Figs.~\ref{Fig2}(d) and (h).
More details regarding the fitting procedure can be found in Appendix~\ref{Fitting}.

%  Characterizing the VL equilibrium state
\subsection{Characterizing the VL equilibrium state}
\label{ESVL}
Several competing factors determines both the structural and dynamic properties of vortex matter.
While the repulsive vortex-vortex interaction favors the formation of an ordered VL, thermal effects and/or pinning to imperfections can lead to disordering.
For example cooling through the superconducting transition in a constant field will produce an uniform vortex density, but can lead to a disordered VL in materials with a strong peak effect.\cite{Yaron:2011tb,MarzialiBermudez:2017vl}
In such cases an annealing is required to remove disorder frozen in close to the upper critical field and obtain an ordered VL with a well-defined diffraction pattern. 
In materials with weak pinning this can be achieved by applying a damped magnetic field oscillation with an initial amplitude of $\sim10\%$ to ``shake'' the vortices free of their pinning sites,
and allow them to find their equilibrium positions.\cite{Levett:2002ba,Levett:2003ug,Gilardi:2004wk}
In contrast shaking can lead to a further disordering of the VL in superconductors with strong pinning.
%A 50~mT initial amplitude of the damped field oscillation corresponds to 10\% of the dc field, which was previously found to be sufficient to achieve an equilibrium VL in low pinning superconductors where disorder was frozen in during a field cooling procedure.\cite{Levett:2003ug,Gilardi:2004wk}

\begin{figure*}
    \includegraphics{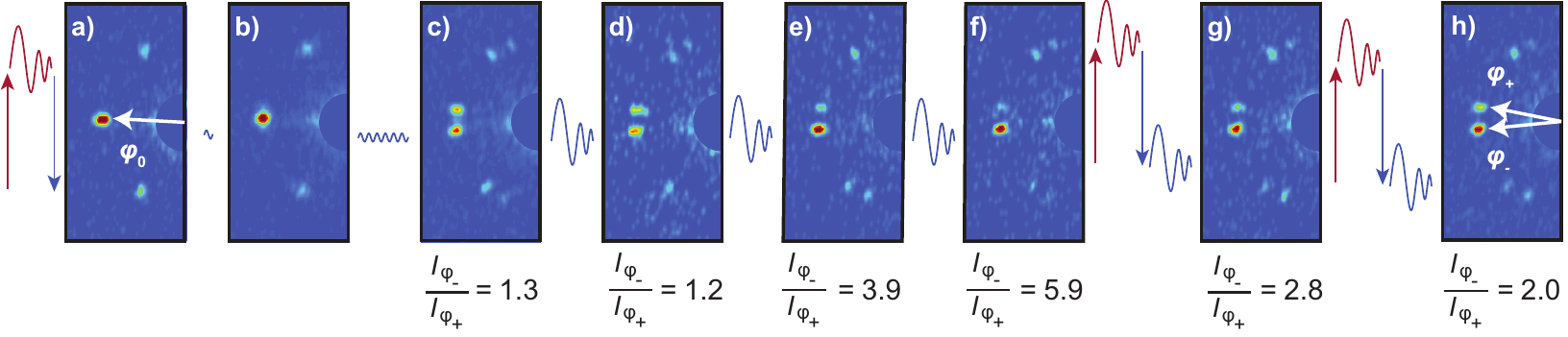}
    \caption{(Color online)
        VL diffraction patterns obtained at 2.6~K following different field/temperature histories.
        (a) F phase.
        (b), (c) Mixed MS/ES following respectively 1 and 786,432 ac field cycles with $\mu_0 \Hac = 0.93$~mT.
        (d), (e), (f) L phase observed after repeated application of a damped field oscillation.
        (g), (h) L phase after a damped field oscillation at 14.2~K, followed by a field oscillation at 2.6~K.
        Each diffraction pattern shows the same region of reciprocal space ($q_x~=[-0.15~,~0]$~nm$^{-1}$, ~$q_y~=~[-0.15~,~0.15]$~nm$^{-1}$). All data is normalized to exposure time and uses the same color scale. Panels (d) to (h) were counted for a shorter length of time, leading to a noisier appearance. %Symbols as defined in Fig.~\ref{Fig1}(e).
        \label{Fig3}}
\end{figure*}
Magnesium diboride exhibits very weak pinning,\cite{Zehetmayer:2002il,Eisterer:2005jf} and always forms an ordered VL with sharp SANS diffraction peaks irrespective of the field/temperature history.\cite{Das:2012cf}
However, as evident from the extensive metastability, in this case a high degree of order does not imply that the vortices are in the ES favored by the vortex-vortex interaction.
Classifying the VL, obtained following the damped magnetic field oscillation with an initial amplitude of 50~mT, as the ES is supported by the diffraction patterns in Fig.~\ref{Fig3}. 
Panels (a) to (c) shows part of the data from Fig.~\ref{Fig2}(d), where the MS F phase is driven towards the ES L phase by successive applications of ac field cycles.
After a single cycle the bulk of the sample remains in the F phase, but some faint L phase intensity is visible at $\varphi = \varphi_-$ and $\varphi_+$ (b).
Additional ac field cycles will drive the VL further towards the L phase, but even after 786,432 cycles there is still some remnant intensity at $\varphi_0$ (c).
A damped field oscillation eliminated the residual F phase intensity, resulting in a L phase throughout the entire sample (d).
Further applications of the same field oscillation without any temperature cycling changes the relative intensity of the two L phase peaks ($I_{\varphi_-}/I_{\varphi_+}$) but not their location at $\varphi - \varphi_0  \approx \pm 7.4^{\circ}$(e,f).
Importantly, we never observe the re-emergence of intensity at $\varphi = \varphi_0$, confirming that the damped field oscillation does indeed drive the VL to the ES.
While it is likely that an initial amplitude of less than 50~mT is sufficient to achieve an ES, this value was used to ensure that the results are unaffected by potential surface barriers for vortex entry and exit. Moreover, the exact value is not expected to affect the main results of this report.

The integrated intensity of the Bragg peaks, obtained from the multi-peak gaussian fits, is proportional to the number of vortices within each of the corresponding VL domain orientations.
While there are most likely many separate domains for both orientations, the total population of each orientation is given by the intensity of the corresponding Bragg peak.
For the particular supercooled case shown in Fig.~\ref{Fig2}(d) and Fig.~\ref{Fig3}(a) to (c), the intensity ratio for the ES domains evolving from the single MS peak is close to unity.
However this is not generally the case, as demonstrated by the different values of the intensity ratio $I_{\varphi_-}/I_{\varphi_+}$ in Fig.~\ref{Fig3}(c) to (h).
This highlights the large degeneracy associated with the L phase, arising from the two possible orientations of individual VL domains.
Successive applications of a damped field oscillation appear to increase the intensity ratio as seen in Fig.~\ref{Fig3}(d) and (f), corresponding to a prevalent F phase domain orientation throughout most of the sample.
Re-preparing a MS F phase at 14.2~K and cooling back to 2.6~K effectively resets the VL, as indicated by the comparatively low intensity ratios in Figs.~\ref{Fig3}(g) and (h).

\begin{figure}
    \includegraphics{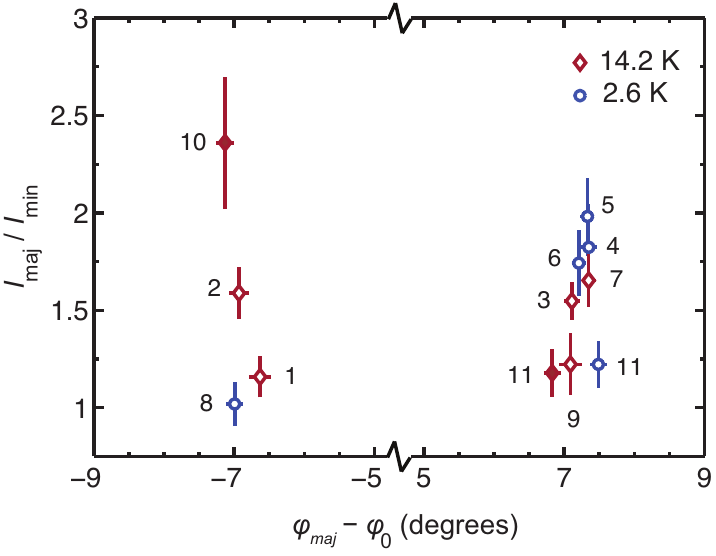} 
    \caption{(Color online)
        Intensity ratio of the VL Bragg peaks for different preparations of the L phase.
        Measurements were performed either on the ES VL at 2.6~K, or on the MS VL after warming to 14.2~K.
        Data points on the left (right) correspond to preparations with the majority domain at a negative (positive) value of $\varphi - \varphi_0$.
        Solid symbols indicate preparations used for measurement sequences discussed in Sect.~\ref{DomPop}.
        \label{Fig4}}
\end{figure}

Figure~\ref{Fig4} shows the relative populations of majority and minority domains, given by the intensity ratio $I_{\text{maj}} / I_{\text{min}}$, for 11 different preparations of the L phase VL.
This demonstrates that the domain population is stochastic in nature,
with intensity ratios as large as 2.4 and a majority orientation corresponding to either the negative or positive $\varphi$ domain in a seemingly random manner.
In contrast to the intensity ratio, the Bragg peak separation is roughly constant $\Delta \varphi \approx 14.2^{\circ}$.
Heating the VL from 2.6~K to 14.2~K does not affect this domain population, as seen for preparation \#11.
For simplicity Fig.~\ref{Fig4} only includes data following an initial damped field oscillation, but as shown in Fig.~\ref{Fig3} (d) to (f) repeated applications will change the intensity ratio.

% L phase domain population
%\subsection{Stochastic nature of the L phase domain population}
%\label{stochastic}

%For the superheated case, the two peaks for the pristine L phase VL in Fig.~\ref{Fig2}(e) have an unequal intensity with a ratio $\simeq$~3:1.
%Repeating the preparation of the MS VL will yield a different relative intensity but always roughly the same peak separation ($\Delta \varphi \approx 14.2^{\circ}$), showing that only the domain population is stochastic in nature.
%The distribution of relative intensities is discussed in more details in Sect.~\ref{stochastic}.

%To prepare the L phase, the sample was first cooled to a temperature of 2.6~K in a magnetic field of 0.5~T.
%Subsequently, a damped dc field oscillation with an initial amplitude of 50 mT was applied to drive the VL to the equilibrium state (ES).
%Prior to initiating measurement sequences like the one shown in Fig.~2(h) of the main text, a characterization of the VL was performed.
%The two Bragg peaks of the L phase correspond to different domain orientations.
%While there are most likely many separate domains for both orientations, the total population of each orientation is given by the intensity of the corresponding Bragg peak.

% L phase domain population
\subsection{Domain population during the metastable to equilibrium state transition}
\label{DomPop}

\begin{figure}
    \includegraphics{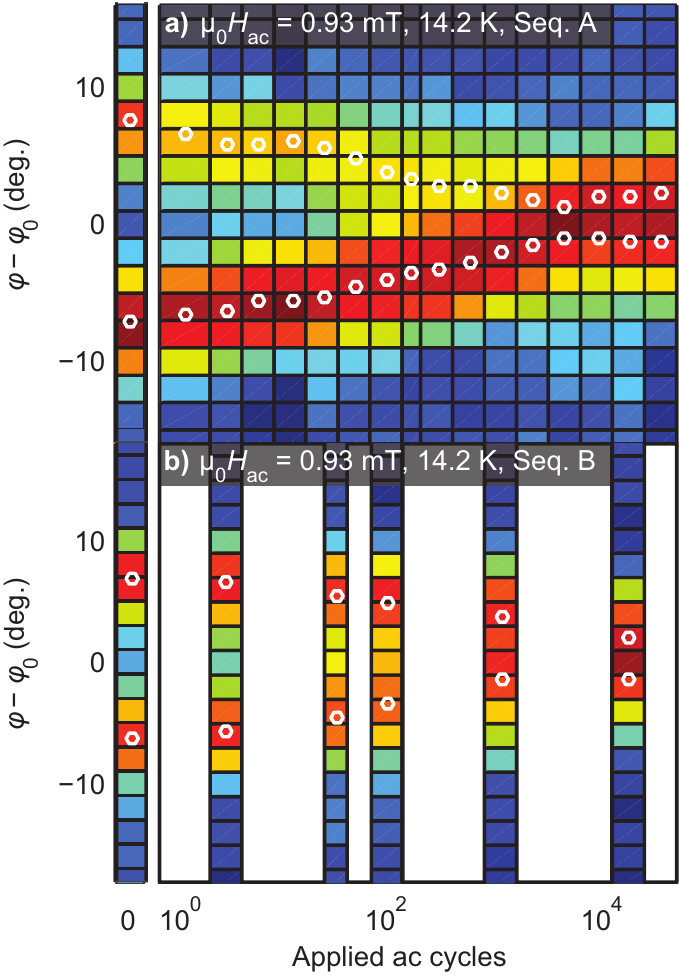}
    \caption{(Color online)
        Azimuthal intensity distribution vs number of ac cycles for two measurements sequences for the superheated L phase with different initial VL domain distributions.
        (a) Sequence~A with $I_{\varphi_-}/I_{\varphi_+} \simeq 2.5$ is the same as in Fig.~\ref{Fig2}(h).
        (b) Sequence~B with  $I_{\varphi_-}/I_{\varphi_+} \simeq 1.2$.
        The left colorbars indicates the pristine MS VL, and the open circles the fitted peak positions.
        \label{Fig5}}
\end{figure}

As discussed above, the relative domain population in the L phase varies from one preparation to the next.
To determine whether this affects the MS to ES transition for the superheated VL, two different measurement sequences with different values of $I_{\varphi_-}/I_{\varphi_+}$ were compared as shown in Fig.~\ref{Fig5}.
Sequence~A (prep.~\#10 in Fig.~\ref{Fig4}) is the same data set as presented in Fig.~\ref{Fig2}(h), with an initial intensity ratio of $\simeq$ 2.5:1. % and Bragg peaks at $\pm 7.3^{\circ}$.
In comparison Seq.~B (prep. \#11) is a less detailed measurement series, with an almost even initial intensity ratio $\simeq$ 1.2:1. % and domain orientations at $\pm 6.6^{\circ}$.
The evolution from the MS to the ES proceeds in a similar manner for both preparations, with the VL domains rotating continuously towards but never fully reaching the ES orientation $\varphi = \varphi_0$.

\begin{figure}
    \includegraphics{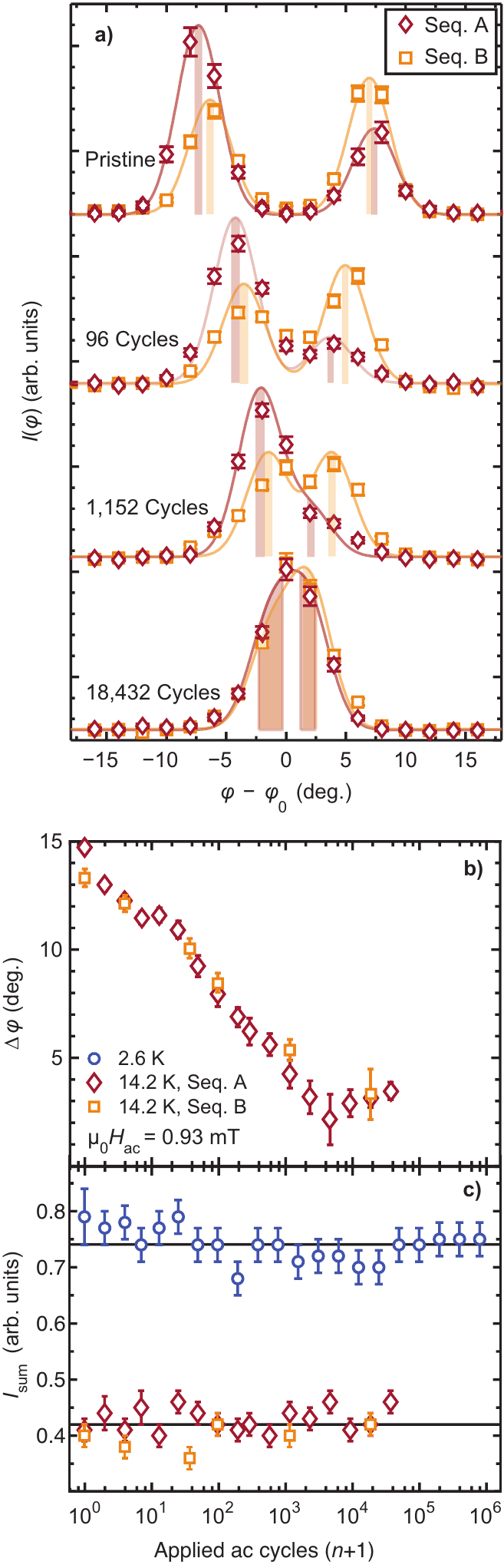}
    \caption{(Color online)
    %Details of the MS to ES transition for the superheated L phase.
    (a) Two-peak gaussian fits to $I(\varphi)$ at four positions along the superheated measurement sequences with $\mu \Hac = 0.93$~mT. Vertical bars indicate the fitted peak centers (position) and uncertainties (width).
    (b) Fitted Bragg peak separation.
    (c) Total measured intensity for the superheated sequences and the supercooled data in Fig.~\ref{Fig2}(d).
    Full lines show the averages.
    \label{Fig6}}
\end{figure}

A more detailed comparison of the azimuthal intensity distribution for the two measurement sequences is given in Fig.~\ref{Fig6}.
Two-peak gaussian fits to the data at several points along the transition  are shown in panel (a).
%Here the azimuthal width was kept equal and constant during fitting for both peaks throughout the transition as described in Appendix~\ref{Fitting}.
The larger errors on the peak positions at the highest cycle count is due to the uncertainty in fitting two maxima when the separation is less than the width.
Figure~\ref{Fig6}(b) shows the VL Bragg peak separation, $\Delta \varphi$, vs the number of applied ac cycles for both measurement sequences. Here the cycle count is offset by one in order to include data for the pristine MS VL. Within measurement error, the results for the two different preparations of the superheated MS VL agree.

The peak separation decreases in a logarithmic fashion up to $\sim 2 \times10^3$ ac cycles, after which it appears to stabilize at $\sim 3^{\circ}$ .
The finite value of $\Delta \varphi$ could in principle be due to a disordered ES, which would lead to a broadening of the azimuthal intensity distribution.
While it is not possible to differentiate between a saturation or a broadening directly from $I(\varphi)$ when $\Delta \varphi$ is small, a significant disordering of the VL would also be expected to lead to a decrease of the scattered intensity.
Figure~\ref{Fig6}(c) shows the total scattered VL Bragg peaks intensities for the superheated measurement sequences, with no observable change throughout the transition from MS to ES.
This makes a disordering of the VL unlikely, and the saturation of $\Delta \varphi$ thus suggests that the ac field amplitude of 0.93~mT is insufficient to completely eliminate all domain boundaries and drive the VL to the global ES.
Similarly, a small residual intensity is seen in Fig.~\ref{Fig2}(c) at $\varphi = \varphi_0$ for the supercooled VL.
In this case the total scattered also remains constant throughout the measurement sequence, Fig.~\ref{Fig6}(c).
The lower scattered intensity for the superheated case is due to the higher measurement temperature.\cite{Eskildsen:2011jp}

\begin{figure}
    \includegraphics{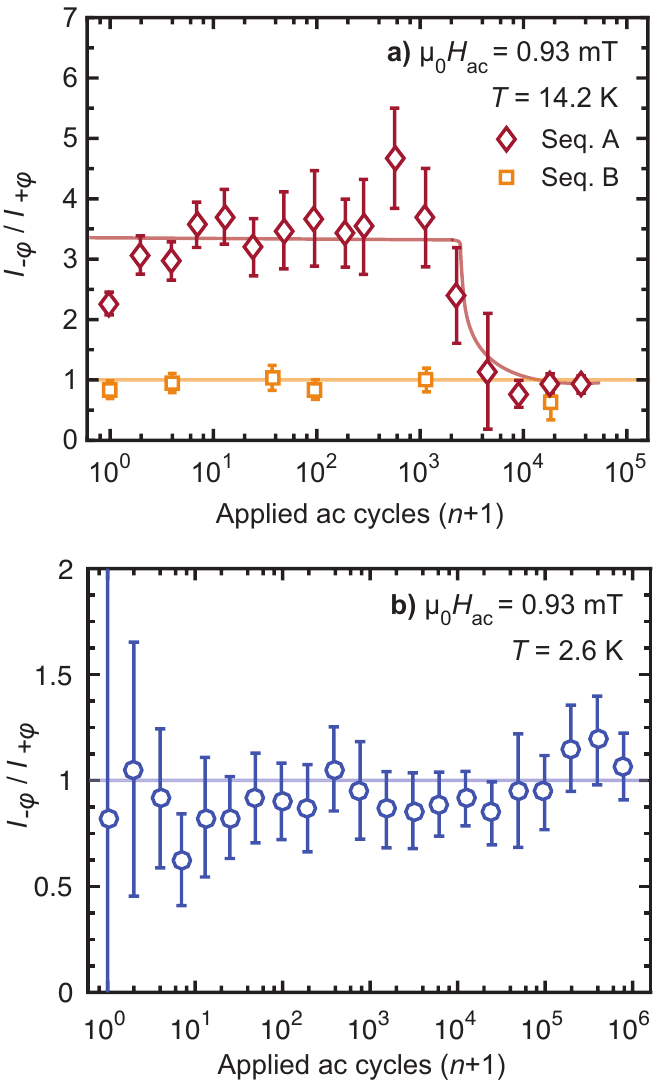} 
    \caption{(Color online)
        Relative domain populations during the MS to ES transition for both the superheated (a) and supercooled (b) case.
        %The shaded area corresponds to the same range of ac field cycles as in Fig. 3 of the main text.
        For the supercooled VL the intensity ratio is for the two Bragg peaks corresponding to ES domains.
        \label{Fig7}}
\end{figure}

Figure~\ref{Fig7}(a) shows the evolution of the intensity ratios corresponding to the two superheated measurement sequences.
For seq.~A, the ratio remains constant ($\simeq$ 3) up to approximately $10^3$ cycles, after which it decreases rapidly towards unity.
This shift towards an equal domain population coincides with the point where no further VL rotation is observed.
In contrast, the ratio for seq.~B does not deviate significantly from unity for the entire measurement sequence.
The larger error bars on $I_{\varphi_-} / I_{\varphi_+}$ before the relaxation towards unity for seq.~A are due to the uncertainty in fitting the two close maxima of different magnitude.
Figure~\ref{Fig7}(b) shows the intensity ratio for the ES VL domains ($\varphi - \varphi_0 \approx \pm 7^{\circ}$), corresponding to the supercooled measurement sequence in Fig.~\ref{Fig2}(a) to(d).
This stays fairly constant and close to unity for the whole measurement sequence.
The error bars decrease as the intensity of the ES domains increases.

\subsection{Transition kinetics and activated behavior}
% Transition kinetics and activated behavior
The presence of metastable phases in {\mgb} cannot be understood based on the single domain free energy shown in Fig.~\ref{Fig1}(d).
Rather, it requires the presence of additional energy barriers to prevent individual VL domains from rotating to the equilibrium orientation.
The absence of any thermally driven relaxation towards the ES within experimental time scales is consistent with the small Ginzburg number $Gi \sim 10^{-6}$ for {\mgb}.\cite{Klein:2010hh}
It also implies that the pristine MS VL will not depend on the heating or cooling rates, nor on the temperature in the F phase where the damped field oscillation was performed for the supercooled case.

Previously we have reported studies of the MS to ES VL transition kinetics for the supercooled case.\cite{Louden:2019bq}
Here the transition was quantified by the remaining metastable volume fraction ($\fms$). %   , defined as the ratio of the intensity of the MS (middle) Bragg peak and the total scattered intensity obtained from three-peak gaussian fits to the azimuthal intensity distribution as shown in Figs.~\ref{Fig2}(a) to \ref{Fig2}(c).
The relaxation towards the ES was modeled as an activated behavior, driven by an increasing number ($n$) of applied ac field cycles, using the an expression for the $\fms$ decay rate given by
\begin{equation}
	\frac{d \fms}{dn} = - \fms \; \exp \left[ -\tilde{H}/\Hac \right].
	\label{Eq:actSC}
\end{equation}
Here the activation field $\tilde{H}$ represents the barrier between MS and ES VL domain orientations and the proportionality to $\fms$ accounts for the remaining metastable volume available for ES domain nucleation and/or growth.
The activation field for the different values of $\Hac$ collapse on a single curve, showing that the ac amplitude and cycle count act as, respectively, the effective ``temperature" and ``time".\cite{Louden:2019bq}

\begin{figure}
	\includegraphics{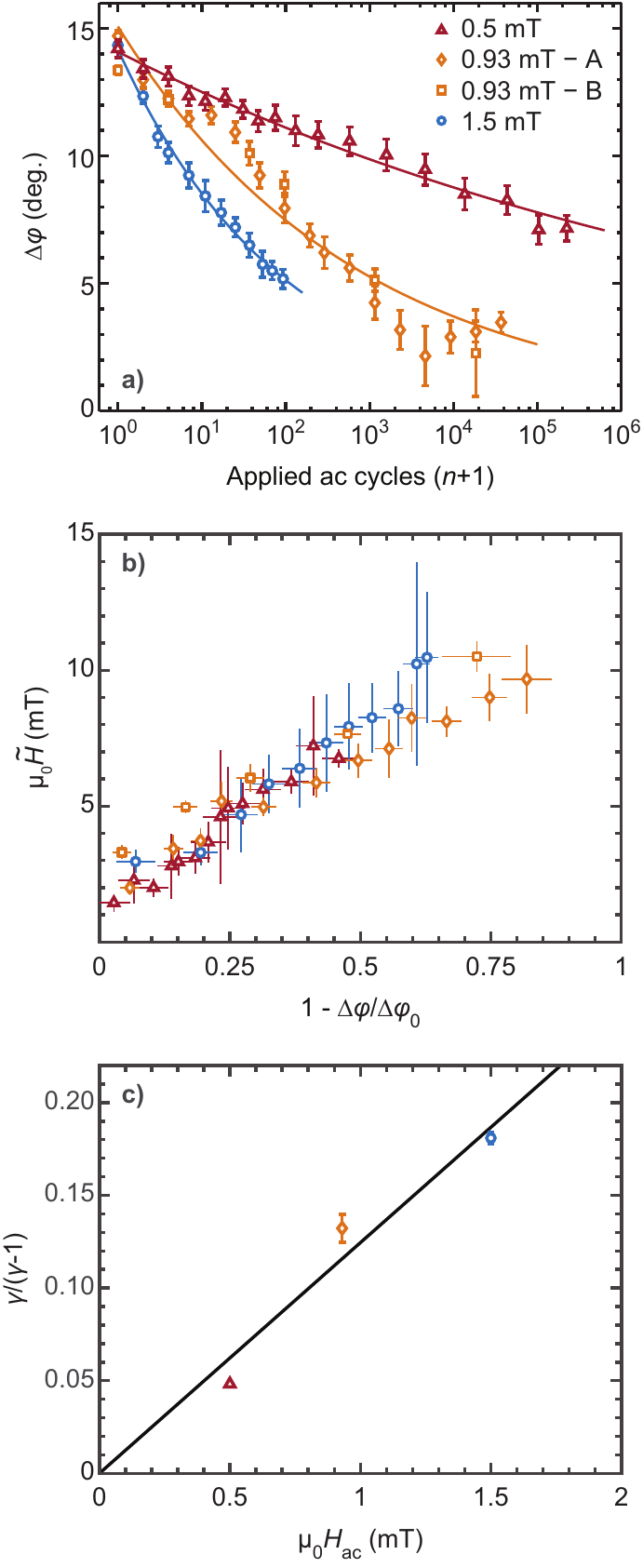}
	\caption{
		(a) Fitted Bragg peak separation as a function of applied cycles for different ac field amplitudes.
		Lines are fits to Eq.~(\ref{Eq:DPhifit}).
		(b) Activation field determined using Eq.~(\ref{Eq:actSH}).
		(c) Relationship between the parameter $\gamma$ in Eq.~(\ref{Eq:Htilde}) and the AC field amplitude, fitted by a straight line through the origin.
		\label{Fig8}}
\end{figure}

In the present work we extend measurements of the VL kinetics to the superheated case.
Figure~\ref{Fig8}(a) shows $\Delta \varphi$ versus $n$ for the three the different ac field amplitudes used.
As was the case for the supercooled MS VL, fewer ac cycles are required to drive the VL towards the ES as $\Hac$ is increased.
However, in the superheated case the transition to the ES is continuous, and the entire VL will continue to rotate until the peak splitting reaches zero.
There is thus no depletion of the VL volume fraction which remains in the MS before the transition is complete.
As a result the transition is modeled by
\begin{equation}
	\frac{d \Delta \varphi}{dn} = - \Delta \varphi_0 \; \exp \left[ -\tilde{H}/\Hac \right],
	\label{Eq:actSH}
\end{equation}
where the prefactor $\Delta \varphi_0$ is the splitting of the pristine MS VL (rather than $\Delta \varphi$).
Figure~\ref{Fig8}(b) shows the activation field obtained from Eq.~(\ref{Eq:actSH}), where each value of $\tilde{H}$ was determined from two adjacent values of $\Delta \varphi(n)$ by
$\tilde{H} = - \Hac \ln \left[ - \tfrac{(\Delta \varphi(n_{i+1}) - \Delta \varphi(n_i))/\Delta \varphi_0}{n_{i+1} - n_i} \right]$.
Within the scatter of the data $\tilde{H}$ collapses onto a single curve, suggesting a near universal behavior consistent with an activated transition.
Like in the supercooled case the activation field increases as the transition towards the ES progresses,\cite{Louden:2019bq} equivalent to an aging of the VL when the ac cycles count is interpreted as an effective ``time''.\cite{Henkel:2010wn}
However, the values of $\tilde{H}$ is twice as large for the superheated case.
We do not currently have an explanation for the difference in activation field, but note that it may be due to the different nature of the transition (continuous vs discontinuous).
Furthermore, in the superheated case each MS VL domain will rotate either clockwise or counterclockwise depending on the sign of $\varphi$ whereas in the supercooled case each VL domain has two decay ``channels'' corresponding to ES domains rotated in opposite directions.

To parameterize the MS to ES transition, the Bragg peak separation was fitted by
\begin{equation}
	\Delta \varphi(n) = \Delta \varphi_0 \; (n+1)^{\gamma},
	\label{Eq:DPhifit}
\end{equation}
which again is in analogy with the functional form used previously for the supercooled case.\cite{Louden:2019bq}
As seen in Fig.~\ref{Fig8}(a), the fits provide an excellent description of the data for all ac field amplitudes and allow a direct calculation of the activation field by
\begin{equation}
	\tilde{H}/\Hac = \ln \left( -\frac{1}{\gamma} \right) + \frac{\gamma - 1}{\gamma} \; \ln \left( \frac{\Delta \varphi_0}{\Delta \varphi} \right).
	\label{Eq:Htilde}
\end{equation}
From this we obtain a non-zero value of $\tilde{H}(n=0) \simeq 2$~mT, which prevents a spontaneous rotation of the MS VL.
In addition, we find $\gamma/(\gamma - 1)$ to be directly proportional to the ac field amplitude, as shown in Fig.~\ref{Fig8}(c).
Consequently, the superheated MS to ES transition is determined by a single parameter, which depends only on $\Hac$.
Once again this is analogous to be previously reported behavior for the supercooled case.\cite{Louden:2019bq}

% DISCUSSION AND SUMMARY
\section{Discussion and Summary}

% Continuous vs discontinuous
The main results of this report is the qualitatively different nature by which the supercooled and superheated VL returns to the equilibrium state, while at the same time exhibiting similar transition kinetics and activated behavior.
In the supercooled case the transition proceeds in a discontinuous manner, with VL domains nucleating at one of the two equilibrium orientations and subsequently growing at the expense of the metastable domains.
This is in striking contrast to the continuous transition observed for the equilibrium VL, where domains gradually rotate away from $\varphi = \varphi_0$ as a function of magnetic field and/or temperature.\cite{Das:2012cf}
Moreover, it is opposite to the common situation where a discontinuous order transition may be broadened by defects or impurities and appear continuous.\cite{Imry:1979wx,Soibel:2001fg,Levett:2002ba}
In contrast, the transition for the superheated case is continuous with domains that rotate towards the ES orientation.
This dichotomy in the MS to ES transition adds to the already unusual behavior of the VL in {\mgb},
where the metastable states are associated with a continuous equilibrium phase transition and therefore not expected to lead to hysteresis.

% Domain boundaries
We speculate that the difference between the two cases arises from the qualitatively different single domain free energy curves shown in Fig.~\ref{Fig1}(d).
For the supercooled case the system is in an unstable equilibrium configuration with $d(\delta F)/d\varphi = 0$, and given the energy costs associated with the creation of domain boundaries a small rotation would likely result in a net increase of the total energy.
Instead, it is more favorable for VL domains to nucleate and grow in the ES orientation.
In comparison the superheated VL is already split into domains rotated in opposite directions.
Moreover, the system is in a true non-equilibrium configuration with $d(\delta F)/d\varphi \neq 0$, where any gradual rotation of the VL domains towards the ES orientation reduces the energy.
%However, although both metastable states are derived from the same continuous phase transition, they do correspond to qualitatively different single domain free energy curves as shown in Fig.~\ref{Fig1}(d), namely an unstable equilibrium in the supercooled case but a true non-equilibrium configuration for the superheated case.
Further experimental evidence for the importance of domain boundaries in stabilizing the metastable VL states comes from the results in Fig.~\ref{Fig6}(b).
Here two different superheated VL configurations with different domain population ratios were found to proceed towards the ES in exactly the same manner.
This shows that it is the presence of domain boundaries, rather than the size or distribution of individual domains, that determines the VL behavior.

% Activation
Since a rotation of the VL involves the rearrangement of a large number of vortices, the existence of energy barriers is not surprising.
The kinetics for the supercooled and superheated VL is similar as shown in Fig.~\ref{Fig8}(b) and Ref.~\onlinecite{Louden:2019bq}, with an activation barrier that increases as the system gets closer to the global equilibrium configuration.
This indicates that in both cases the behavior of the VL is likely governed by the same mechanism, despite the different nature of the MS-to-ES transition.
We also note that an increasing activation energy has also been found for isothermal martensitic phase transformations in maraging steel, where domain formation and the motion of domain boundaries are known to play a crucial role.\cite{Martin:2010mse}

% Analogies to SkL
Given the many similarities between vortices and skyrmions, it is possible that the effects reported here for the VL will also occur for the skyrmion lattice (SkL).
Recently, a similar scenario was proposed to explain hysteretic behavior and slow dynamics at a nominally continuous transition of the helimagnetic order in MnSi.\cite{Bauer:2016wp}
Discontinuous reorientation transitions have also been observed for the SkL, and metastable configurations have been achieved by different field/temperature histories\cite{Makino:2017hh} or by rotating the sample in a constant magnetic field.\cite{Bannenberg:2017ws}
Moreover, an applied dc electric field can change the preferred orientation of the SkL, which can be driven to the new equilibrium state in a continuous manner by the application of ac magnetic field cycles.\cite{White:2014ji}
Finally, the skyrmion lattice (SkL) can be thermally quenched far below the equilibrium phase, which expands the range of fields where skyrmions are stable and may induce a symmetry transition from a hexagonal to a square lattice.\cite{Nakajima:2017uc,Karube:2016bs}
%Given the parallels between skyrmions and vortices, it is unsurprising that robust metastable states are observed in both cases and we expect that they will belong to the same universality class.
%
%The metastability can not be understood from the single domain free energy in Fig.~\ref{Fig1}(d).
%Moreover, VL domains were previously found to persist in the metastable orientation in the presence of vortex motion induced by a change in the applied field.\cite{Rastovski:2013ff}
%This excludes vortex pinning to impurities as an explanation for the metastability, implying a scenario where boundaries between VL domains create energy barriers which prevent them from rotating to their equilibrium orientation relative to the host crystal.
%Given that a rotation of the VL involves the rearrangement of a large number of vortices, the existence of such energy barriers would not be surprising.
%Recently, a similar scenario was proposed to explain hysteretic behavior and slow dynamics at a nominally continuous transition of the helimagnetic order in MnSi.\cite{Bauer:2016wp}
%Experimentally, the results in Fig.~\ref{Fig3}(a) provide further support for the importance of domain boundaries in stabilizing the metastable VL states in {\mgb}.
%Here two different superheated VL configurations with different domain population ratios were found to proceed towards the ES in exactly the same manner.
%This shows that it is the presence of domain boundaries, rather than the size or distribution of individual domains, that determines the VL behavior.

% Glases
Finally we note that glasses are the quintessential example of a supercooled, metastable configuration observed in conjunction with a thermally driven transition.
Similarities between the metastable VL states and supercooled liquids and other structural glasses includes an activated transitions between states resulting from a complicated energy landscape, and a behavior that is governed by domains and domain walls.\cite{Kirkpatrick:1989ii,Lubchenko:2007cc,Nussinov:2017fm}
Further support for this analogy comes from the slowing kinetics (aging) in Fig.~\ref{Fig8}(b).
Here it is important to acknowledge that describing the MS VL as ``supercooled" or ``superheated" is strictly speaking incorrect, as thermal excitations are too weak to affect the vortices in {\mgb}.
We have nonetheless used this nomenclature since it provides an intuitive and convenient way to discuss our measurements and results.  
Further, there is, as already mentioned, a straight forward analogy between the current situation and an ordinary thermally driven transition since the ac field amplitude determines the magnitude of the vortex motion,
and may therefore be interpreted as an effective ``temperature''.
Similarly the number of ac cycles correspond to an effective ``time''.
In summary, we have performed detailed measurements of non-equilibrium VL phase transitions in {\mgb}.
We studied how metastable VLs, obtained by either supercooling or superheating across an intrinsically continuous phase transition, return to the ES under the influence of an ac magnetic field.
In the case of the supercooled VL, this occurs in a discontinuous manner.
In contrast, the transition takes on a continuous nature for the superheated case.
We suggest that this qualitative difference is due to being respectively in an unstable equilibrium or a true non-equilibrium single domain configuration.
Despite the different nature of the transition for the two cases the kinetics are similar, with an activation barrier that increases as the system approaches the equilibrium configuration.
Our results provide further evidence that domain boundaries are responsible for the metastable VL states.
Additional studies to provide real space information about the VL, either experimentally (e.g. by STM) or by non-equilibrium molecular dynamics simulations,\cite{Olszewski:2018fp} would be a valuable complement to our SANS results.
To our knowledge there has been only a single theoretical study of VL domain boundaries,\cite{Deutsch:2010bi} and more work is needed to fully understand our observations.

% ACKNOWLEDGEMENTS
\section*{Acknowledgements}
We are grateful to J.~Karpinski for providing the {\mgb} single crystal used for this work.
We acknowledge useful discussions with E.~M.~Forgan, B.~Janko, K.~Newman, M.~Pleimlimg, and U.~C.~T\"{a}uber,
and assistance with the SANS experiments and data analysis from J.~Archer, S.~J.~Kuhn, and A.~Leishman.
This work was supported by the U.S. Department of Energy, Office of Basic Energy Sciences, under Award No. DE-SC0005051.
A portion of this research used resources at the High Flux Isotope Reactor, a DOE Office of Science User Facility operated by the Oak Ridge National Laboratory.

% BEGIN APPENDICES 
\appendix

\section{Fitting algorithm}
\label{Fitting}
%As described in the main text and evidenced by the colormaps of Fig.~2 therein, the supercooled and superheated VLs followed different transition pathways to the ES.
%This necessitates a different data analysis as discussed in the following.
The azimuthal intensity distribution in both the supercooled and superheated case is well described by a sum of gaussians, corresponding to the Bragg peak for each of the VL domain orientations:
\begin{equation}
  	I (\varphi) = I_0 + \sum\limits_{j = 1}^{\text{\# peaks}} \frac{I_j}{w_j} \; \exp \left[ - 2 \sqrt{\log \, 4} \left( \frac{\varphi - \varphi_{j}}{w_j} \right)^2 \right].
	\label{Eq:Ifit}
\end{equation}
Here $I_0$ is a constant accounting for isotropic background scattering,  $I_j$ is the integrated intensity, $w_j$ is the full width half maximum (FWHM), and  $\varphi_{j}$ is the center for the $j$th Bragg peak. 
The individual peak intensities ($I_j$) are  proportional to the number of scatterers in the corresponding domain orientation.
All supercooled VL transition data were fit with three gaussian peaks, while all superheated data were fit with two.

\begin{figure*}
    \includegraphics{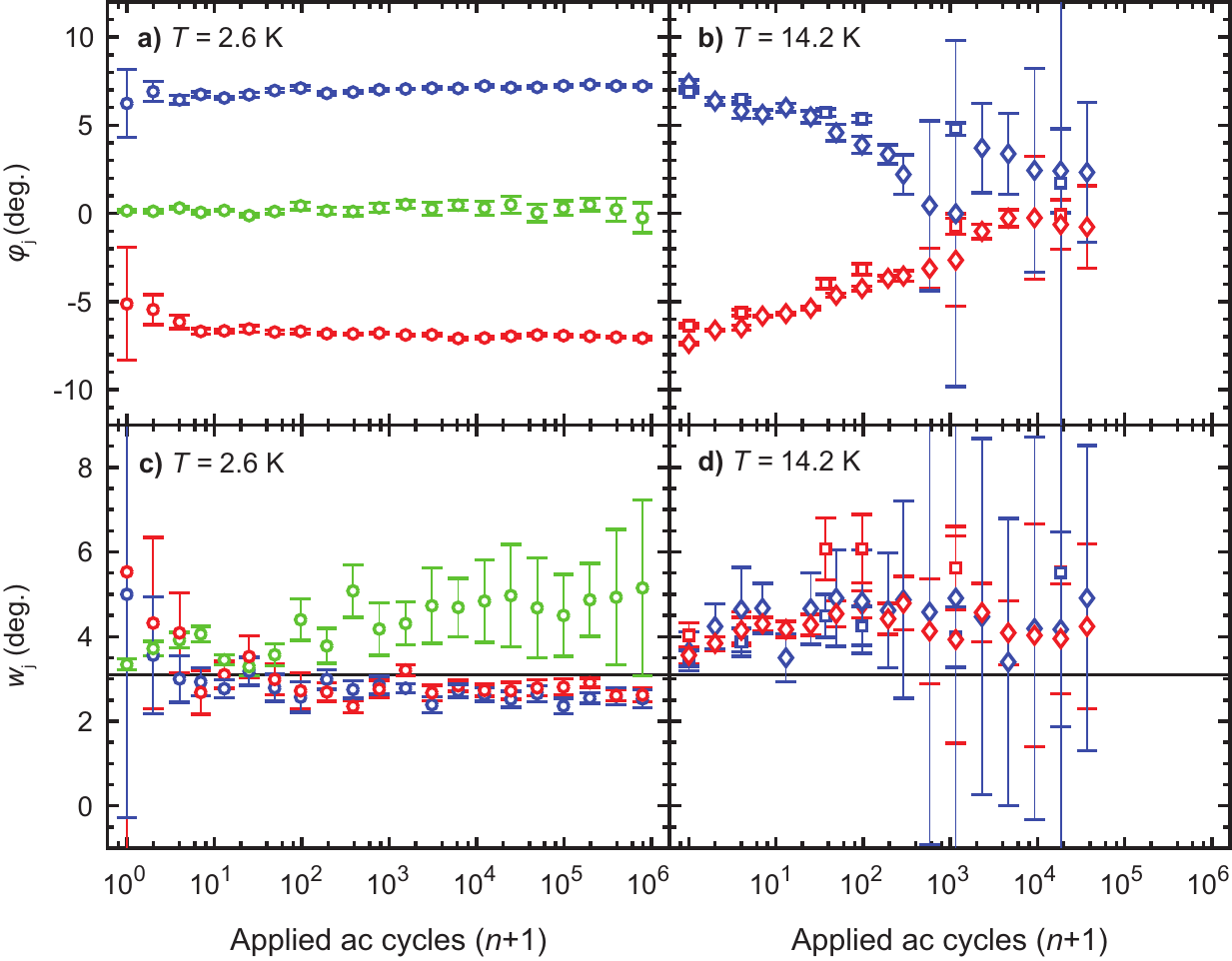}
    \caption{
        Results of unconstrained fits to Eq.~(\ref{Eq:Ifit}) for the measurement sequences in Figs.~\ref{Fig2}(d) and \ref{Fig2}(h).
        (a),(b) Fitted peak centers.
        (c),(d) Fitted peak widths.
        The angular resolution ($w_{\text{res}} = 3.1^{\circ}$) is shown by the black line.
  	    %The number of ac cycles were offset by one in order to display the pristine MS ($n = 0$).
	    \label{Fig9}}
\end{figure*}

Figure~\ref{Fig9} shows the centers and widths obtained from fits to Eq. \eqref{Eq:Ifit}.
In the supercooled case, the Bragg peak intensity is transferred from the MS domains (green) to the ES domains (red, blue).
Once the Bragg peaks for the ES domains are well developed ($n \geq 10^1$) the fitted widths are essentially resolution limited, and consistently smaller than those of the metastable domains, Fig.~\ref{Fig9}(c).
While more ordered ES domains (narrower peaks) would not be surprising, it is not possible to make a definitive conclusion in this regard within the precision of the fits.
We note that error on the fitted peaks centers and widths are greater for the low intensity peaks, such as the ES domains early in the supercooled sequence or the MS domains after $\sim 10^2$~ac cycles.
In the superheated case, it is difficult to de-convolute the effects of the domain rotation from a potential broadening due to VL disordering.
Furthermore, once the peaks have merged near the end of the transition ($n \geq 10^2$) it is not possible to resolve two separate gaussians, leading to a large increase in the errors on the fitted positions and widths in Fig.~\ref{Fig9}(b) and (d).
In summary, the data in Fig.~\ref{Fig9} does not indicate that the ac cycles cause a significant disordering of the VL.
This conclusion is supported by the constant total scattered intensity for both the supercooled and superheated measurement sequences shown in Fig.~\ref{Fig6}(c).
This is in contrast to the systematic decrease as the number of ac cycles is increased, which one would expect in the case of a VL disordering.

\begin{figure*}
    \includegraphics{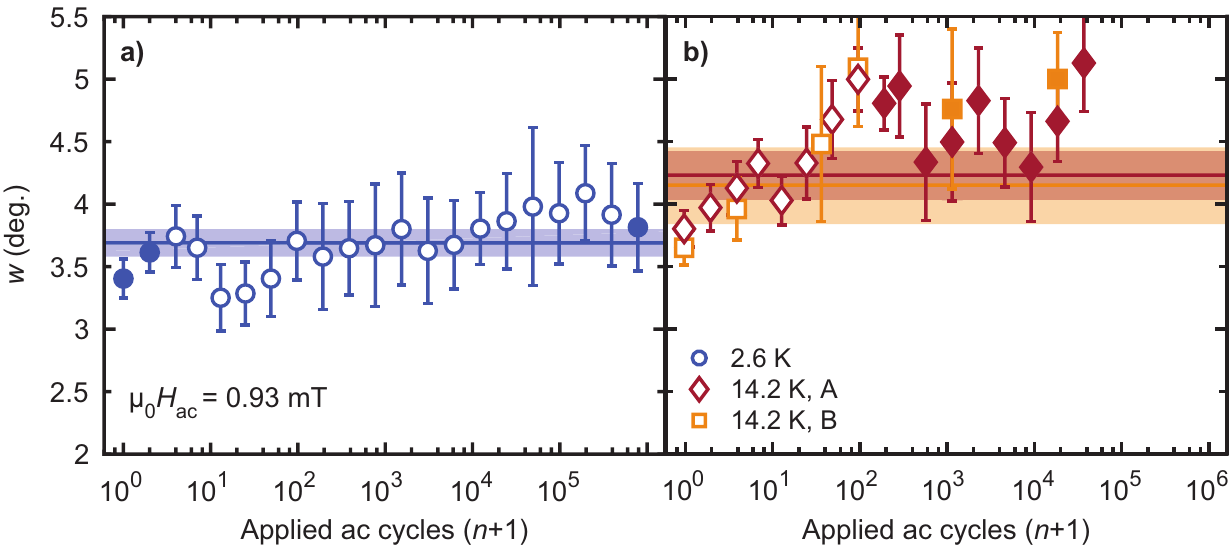}
    \caption{Determination of the average Bragg peak widths (FWHM).
		Fits were performed using Eq.~\eqref{Eq:Ifit}, with all peaks constrained to have the same width.
		Results shown by open symbols met the criteria described in the text and were included in the determination of the final average width (solid line) and standard deviation (shaded area).
  		%ac cycles have been offset by one in order to display the pristine MS ($n = 0$).
        \label{Fig10}}
\end{figure*}

Based on the above discussion we have constrained the fit of the SANS data, using a constant width ($w_j \equiv w$) for all the VL Bragg peaks throughout a given measurement sequence.
While this eliminates potential information contained in the peak widths, it leads to a more precise determination of the peak positions and intensities.
The following protocol was used to determine the most appropriate width to use for a given sequence.
First, initial estimates for the peak positions and widths were determined from the clearly resolved peaks in the first  ($n = 0$) and last ($n = n_{\text{max}}$) measurements.
These values were used to seed independent fits of the individual SANS measurements in the sequence, using a common width for all the peaks.
The resulting widths are shown in Fig. \ref{Fig10}.
%\begin{figure}
%  \includegraphics{FigS7.eps}
%  \caption{Determination of the average Bragg peak widths (FWHM).
%		Fits were performed using Eq.~\eqref{eq1}, with all peaks constrained to have the same width.
%		Only results shown by open symbols met the criteria described in the text and were included in the determination of the final average width (solid line) and standard deviation (shaded area).
%  		ac cycles have been offset by one in order to display the pristine MS ($n = 0$).
%                \label{FigS7}}
%\end{figure}
For the final fit of the data the width was fixed to the average value, calculated from peaks where $I_j  / I_{\text{tot}} > 10\%$ for the supercooled case and $\Delta \varphi > 1.75 w$ for the superheated case.
Data that meet these criteria are shown by the open symbols in Fig.~\ref{Fig10}.

%\newpage
% SUPPLEMENTAL MATERIALS
%\section*{Supplemental Material}
%In the following we present additional background information, data, and analysis details related to our small-angle neutron scattering (SANS) studies of the metastable vortex lattice (VL) states in {\mgb}.
%This includes a discussion of how the equilibrium state (ES) is determined, an additional measurement sequence for the superheated case, movies illustrating how the supercooled and superheated metastable VL transition to the ES state under the influence of ac magnetic field cycles, and details concerning the data analysis and fitting procedure.

%\subsection*{Movies}
%\subsection*{Movie S1}
%Supercooled F phase undergoing a discontinuous transition to the ES L phase at 2.6~K.
%The inset shows the number of ac field cycles ($\mu_0 \Hac = 0.93$~mT, 250~Hz).
%This corresponds to the data shown in Figs.~2(d) of the main text.

%\subsection*{Movie S2}
%Superheated L phase undergoing a continuous transition to the ES F phase at 14.2~K.
%The inset shows the number of ac field cycles ($\mu_0 \Hac = 0.93$~mT, 250~Hz).
%This corresponds to the data shown in Figs.~2(h) of the main text.

%\subsection*{Movie S3}
%Superheated L phase undergoing a continuous transition to the ES F phase at 14.2~K.
%The inset shows the number of ac field cycles ($\mu_0 \Hac = 0.93$~mT, 250~Hz).
%This corresponds to the data shown in Fig.~5(b) of the main text.

%\newpage

% BIBLIOGRAPHY
\newpage %\vspace{1cm} \newpage 
\section*{References}
%\bibliography{MgB2NonEq}


\newpage

\end{document}